\documentclass[conference]{IEEEtran}
\ifCLASSINFOpdf
\else
\fi
\usepackage{epsfig,makeidx,color}
\usepackage{graphicx}
\usepackage{amsbsy}
\usepackage{amsmath}
\usepackage{amssymb}
\usepackage{euscript}

\newtheorem{mytheorem}{\bf Theorem}

\newtheorem{corollary}{\bf Corollary}

\newtheorem{myremark}{Remark}

\newtheorem{myresult}{\bf Result}

\newcommand {\Define} {\stackrel {\Delta} {=}  }

\begin{document}
%
\title{On the Energy-Spectral Efficiency Trade-off of the MRC Receiver in Massive MIMO Systems with Transceiver Power Consumption}

\author{\IEEEauthorblockN{Sudarshan Mukherjee and Saif Khan Mohammed}
\IEEEauthorblockA{Department of Electrical Engineering, Indian Institute of Technology (I.I.T.) Delhi, India.\\
}
}

\maketitle

\begin{abstract}
We consider the uplink of a multiuser massive MIMO system wherein a base station (BS) having $M$ antennas communicates coherently with $K$ single antenna user terminals (UTs). We study the energy efficiency of this system while taking the transceiver power consumption at the UTs and the BS into consideration. For a given spectral efficiency $R$ and fixed transceiver power consumption parameters, we propose and analyze the problem of maximizing the energy efficiency as a function of $(M,K)$. For the maximum ratio combining (MRC) detector at the BS we show that with increasing $R$, $(M,K)$ can be adaptively increased in such a way that the energy efficiency converges to a positive constant as $R \rightarrow \infty$ ($(M,K)$ is increased in such a way that a constant per-user spectral efficiency $R/K$ is maintained). This is in contrast to
the fixed $(M,K)$ scenario where the energy efficiency is known to converge to zero as $R \rightarrow \infty$.
We also observe that for large $R$, the optimal $(M,K)$ maximizing the energy efficiency is such that, the
total power consumed by the power amplifiers (PA) in all the $K$ UTs is a small fraction of the total system power consumption.
\end{abstract}

\IEEEpeerreviewmaketitle

\section{Introduction}
{\renewcommand{\thefootnote}{} \footnote{Saif Khan Mohammed is also associated with the Bharti School of Telecommunication Technology and Management (BSTTM), I.I.T. Delhi. This work was supported by the
    Extra-Mural Research Grant from the Science and Engineering Research Board (SERB), Department of Science and Technology (DST), Government of India.
    }}
Massive MIMO Systems/Large MIMO Systems/Large Scale Antenna Systems collectively refer to a communication system where a base station (BS) (having several tens to hundred antennas) communicates coherently with a few tens of users on the same time-frequency resource \cite{SPM-paper}, \cite{ELLM}.
Recently massive MIMO Systems have been shown to achieve very high spectral efficiency\footnote{\footnotesize{Throughout this paper, by spectral efficiency we refer to the sum of the spectral efficiencies of all the users.}} and energy efficiency\footnote{\footnotesize{Energy efficiency (bits/Joule) is defined as the average number of bits that are reliably communicated for every Joule of energy spent.}} \cite{TM}, \cite{HNQ}.
Currently there is also a lot of emphasis on energy efficient communication systems \cite{Li}.

In the previous work done on studying the energy versus spectral efficiency trade-off (uplink) of low complexity receivers in massive MIMO systems, it has been assumed that the only power consumed in the system is due to the power radiated by the user terminals (UTs) \cite{HNQ}. In \cite{HNQ}, it has been shown that with perfect channel state information (CSI), for a given spectral efficiency the energy efficiency can be increased in an unbounded manner by increasing the number of BS antennas ($M$) and the number of users ($K$). Increasing $M$ will increase the array gain at the BS, and therefore to achieve a fixed spectral efficiency the required power to be radiated from the UTs will reduce. Similarly, increasing $K$ will reduce the per-user information rate, which then reduces the required power to be radiated by each UT. However in practice,  the transceiver circuits in the UTs and at the BS consume power, which will increase with increasing $M$ and $K$. Therefore, if transceiver power consumption is also taken into account, it is clear that for a given spectral efficiency the energy efficiency will not increase in an unbounded manner with increasing $M$ and $K$.
Motivated by the arguments above, in our recent paper \cite{prev_work}, we had studied
the energy-spectral efficiency trade-off of the Zero-Forcing (ZF) receiver while taking transceiver power consumption into consideration.

In this paper, we extend our work in \cite{prev_work} to the study of the energy-spectral efficiency trade-off for the Maximum Ratio Combining (MRC) receiver, which is known to have an even lower complexity than the ZF receiver (since MRC does not require channel inversion) and also achieves near-optimal performance in massive MIMO systems \cite{TM}, \cite{HNQ}. In Section \ref{Sysmodel}, we explain the system model, define the power consumption parameters taken into consideration, and also state the optimization problem of maximizing the energy efficiency with respect to (w.r.t.) $(M,K)$ for a given spectral efficiency $R$. The optimized energy efficiency is hereby referred to as the ``optimal energy efficiency'' for the given $R$. In this paper we focus our study to the regime where the desired spectral efficiency $R$ is large.\footnote{\footnotesize{We consider a given $R$ to be large enough if the corresponding optimal number of BS antennas and the optimal number of users are much larger than one.}}

The optimal energy efficiency is analyzed in Section \ref{largeR_sec}. 
Analysis in Section \ref{largeR_sec} reveals that, i) for the MRC receiver it is possible to increase $(M,K)$ with increasing $R$ so that the energy efficiency converges to a positive constant as $R \rightarrow \infty$ (while maintaining a constant per-user spectral efficiency $R/K$) (see Corollary \ref{cor_largeRR} and Remark \ref{remark_77}), ii) for sufficiently large $R$ the ZF receiver has a higher optimal energy efficiency than the MRC receiver, iii) for a given finite $R$ the optimal $(M,K)$ which maximize the energy efficiency are finite, and iv) for a given $R$ the optimal energy efficiency reduces with increasing values of the power consumption parameters. Numerical simulation is used to confirm the analysis in Section \ref{largeR_sec}. Simulation results are discussed in Section \ref{sec_sim}. We also observe that for large $R$, the optimal $(M,K)$ maximizing the energy efficiency of the MRC receiver is such that, the total power consumed by the power amplifiers (PA) in all the $K$ UTs is a small fraction of the total system power consumption.

Few other works studying the maximization of energy efficiency of massive MIMO systems with transceiver power consumption have recently appeared in \cite{DHa}, \cite{Emil}. However, in both \cite{DHa} and \cite{Emil}, the authors have not studied the maximization of the energy efficiency jointly w.r.t. $(M,K)$ for a {\em given spectral efficiency} and therefore the optimal energy-spectral efficiency curve is not known, and hence its large $R$ behaviour is also not clear. Addressing this issue, in this paper we analyze the optimal energy-spectral efficiency trade-off curve of the MRC receiver in the large $R$ regime.
 
\section{System model}\label{Sysmodel}
Consider the uplink of a multi-user massive MIMO system where a
BS having $M$ antennas communicates with $K$ single antenna user terminals (UTs).
Let $x_k$ be the complex information symbol transmitted from the $k$-th user.\footnote{\footnotesize{In this paper, we consider the
discrete-time complex baseband equivalent model of the original band-limited passband channel.}} The signal received at the $m$-th BS antenna is then given by
\begin{eqnarray}
\label{sysmodel}
y_m & = & \sum_{k=1}^K \, h_{k,m} x_k  \, + \, n_m  \,\,\,,\,\,\, m=1,2,\cdots,M
\end{eqnarray}
where $n_m$ is the additive white complex circular symmetric Gaussian noise (AWGN) at the $m$-th receiver,
having zero mean and variance $\sigma^2 = N_0 B$.
Here $B$ is the channel bandwidth (Hz), and $N_0$ Watts/Hz is the power spectral density of the AWGN.
Here $h_{k,m} = \sqrt{G_c} g_{k,m} \in {\mathbb C}$ denotes the complex channel gain between the
$k$-th UT and the $m$-th BS antenna. Also, $g_{k,m}, k=1,2,\cdots,K, m =1,2,\cdots,M$ are i.i.d. ${\mathcal C}{\mathcal N}(0,1)$ (circular symmetric complex Gaussian
having zero mean and unit variance). Further, $\sqrt{G_c} > 0 $ models the geometric attenuation and shadow fading, and is assumed to be constant over many coherence intervals and known a priori to the BS.\footnote{\footnotesize{We consider a simple model where the attenuation of each user's signal is the same. This is done so as to study the effects of transceiver power consumption on the energy-spectral efficiency trade-off in a standalone manner.
Incorporating different attenuation factors makes it difficult to analyze and draw basic insights about this trade-off.}} The model in (\ref{sysmodel}) is also applicable to wide-band channels where OFDM is used.

Let the average power radiated from each UT be $p_u$ Watts. The average power consumed by each user's transmitter can then be modeled as $p_{tx} = \alpha p_u \, + \, p_t$ where $\alpha > 1$ models the efficiency of the power amplifier (PA) and $p_t$ is the power consumed by the other signal processing circuits (except the PA) inside the transmitter (e.g., oscillator, digital-to-analog converter, filters) \cite{cmos_design,e_eff_tx}. The power consumed by the PA in each UT is $\alpha p_u$.
Further, let $p_r$ (in Watts) be the average power consumed for signal processing in each BS receiver antenna unit (e.g., per-antenna RF and baseband hardware). The average power consumed at the BS for per-user processing is modeled as $p_{dec}$ (e.g., signal processing of each user's coded information stream, decoding the channel code for each user). Let $p_s$ model any other residual power consumption at the BS which is independent of the number of BS antennas and the number of users.\footnote{\footnotesize{This can essentially incorporate any other source of power consumption which is independent of $M$ and $K$.}} 
Then the total system power consumed is
\begin{eqnarray}
\label{P_eqn0}
P & = & K p_{tx} \, + \, \overbrace{(K p_{dec}  + M p_{r} + p_s)}^{\mbox{\small{Power consumed at BS}}}  \nonumber \\
& = & K (\alpha p_u + p_t  + p_{dec}) \, + \, M p_r \, + \, p_s.
\end{eqnarray}
Note that $p_t$ and $p_{dec}$ contribute to $P$ only through their sum
and therefore for brevity of notation, let
\begin{eqnarray}
\label{P_eqn}
p_d & \Define & p_t \, + \, p_{dec} \,\,\,,\,\,\, \mbox{and therefore} \nonumber \\
P & = & K ( \alpha p_u + p_d  ) + M p_{r} + p_s.
\end{eqnarray}
The energy efficiency (bits/Joule) is given by\footnote{\footnotesize{In one second, $RB$ bits are communicated and the total power consumption is $P$ Joules.}}
\begin{eqnarray}
\label{e_eqn}
\eta & = & \frac{ R B }{P},
\end{eqnarray}
where $R$ is the spectral efficiency in bits/s/Hz.
Multiplying (\ref{P_eqn}) on both sides by $G_c/N_0 B$ and then using (\ref{e_eqn})
on the left hand side (L.H.S.) we get
\begin{eqnarray}
\label{eqn_2}
\frac{G_c R}{ N_0 \eta}  & = &  \alpha  K \, \frac{G_c p_u }{N_0 B}   \, + \, K \frac{G_c p_d }{N_0 B} \, + \, M \frac{ G_c  p_r }{ N_0 B} \, + \, \frac{ G_c  p_s }{ N_0 B}.
\end{eqnarray}
For brevity of notation, we make the following definitions\footnote{\footnotesize{Studies have shown that the power consumption in band-limited wireless transceiver circuits is typically proportional to $N_0 B$ (the constant of proportionality usually depends on technology and design parameters) \cite{Nossek},\cite{UDMCMOS}.}}
\begin{eqnarray}
\label{defs_eqn}
\zeta  \Define  \frac{\eta N_0 } {G_c} \,\,\,,\,\,\,
\rho_r  \Define  \frac{G_c p_r }{N_0 B} \,\,\,,\,\,\, \rho_d  \Define  \frac{G_c p_d }{N_0 B } \nonumber \\
\rho_s  \Define  \frac{G_c p_s }{N_0 B } \,\,\,,\,\,\, \gamma \Define \frac{G_c p_u }{N_0 B}.
\end{eqnarray}
Since $\zeta$ depends on the system parameters $\Theta \Define (R, \alpha, \rho_r, \rho_d, \rho_s)$ and $(M,K)$, we subsequently use the notation $\zeta(M, K,\Theta)$ to highlight this dependence.
Using (\ref{defs_eqn}) in (\ref{eqn_2}) we get
\begin{eqnarray}
\label{eqn_12}
\frac{R}{\zeta(M, K , \Theta)}
& \hspace{-3mm} = & {\Big (}   \alpha K \gamma + ( K \rho_d + M \rho_r + \rho_s)  {\Big )}.
\end{eqnarray}
Note that on the right hand side (R.H.S.) of (\ref{eqn_12}), the first term corresponds to power consumed by the PAs in the $K$ UTs, whereas the second term $(M \rho_r + K \rho_d + \rho_s)$ corresponds to the power consumed at the BS and the transmitter circuitry in the UTs.

In a massive MIMO MRC receiver with perfect CSI, an achievable spectral efficiency is given by \cite{HNQ}
\begin{eqnarray}
R  & = & K \, \log_2{\Big ( } 1 \, + \, \frac{G_c p_u (M - 1)}{G_c p_u (K -1) \, + \, N_0 B}  {\Big )}  \nonumber \\
& = & K  \, \log_2 {\Big (} 1  \, + \, \frac{\gamma_{_{\mbox{\footnotesize mrc}}} (M - 1)}{\gamma_{_{\mbox{\footnotesize mrc}}} (K -1) \, + \, 1} {\Big )}  \nonumber \\
\gamma_{_{\mbox{\footnotesize mrc}}} & \Define & \frac{G_c p_u }{N_0 B}
\end{eqnarray}
from which it follows that
\begin{eqnarray}
\label{gamma_expr_mrc}
\gamma_{_{\mbox{\footnotesize mrc}}} & = & \frac{2^{\frac{R}{K}} - 1}{ (M - 1) \, - \, (K - 1) (2^{\frac{R}{K}} - 1)}
\end{eqnarray}
Since $p_u > 0$, $\gamma_{_{\mbox{\footnotesize mrc}}}  > 0$ and therefore
for a given $R$, $(M,K)$ must belong to the set
\begin{eqnarray}
\label{zrset_def}
{\mathcal Z}_{_R} & = & {\Big \{ } (M , K)  \,|\, M, K \in {\mathbb Z} \,,\, K \geq 1 \,,\, \nonumber \\
& & \hspace{10mm}   (M - 1) > (K - 1) (2^{R/K} - 1)  {\Big \}}.
\end{eqnarray}
In a massive MIMO ZF receiver with perfect CSI, an achievable spectral efficiency is given by \cite{HNQ}
\begin{eqnarray}
R  & = & K \, \log_2{\Big ( } 1 \, + \, \frac{G_c p_u (M - K)}{N_0 B}  {\Big )}  \nonumber \\
& = & K  \, \log_2 {\Big (} 1  \, + \,  (M - K) \gamma_{_{\mbox{\footnotesize zf}}}{\Big )}
\end{eqnarray}
from where it follows that
\begin{eqnarray}
\label{gamma_expr_zf}
\gamma_{_{\mbox{\footnotesize zf}}} & = & \frac{2^{\frac{R}{K}} - 1}{ M - K}.
\end{eqnarray}
Since $\gamma_{_{\mbox{\footnotesize zf}}}  > 0$ it follows that
$(M,K)$ must satisfy $M > K$.
Using the expressions of $\gamma$ for the MRC and the ZF receivers from (\ref{gamma_expr_mrc}) and (\ref{gamma_expr_zf}) in (\ref{eqn_12}), the energy efficiency with these two different receivers (denoted by $\zeta_{_{\mbox{\footnotesize mrc}}}(M, K , \Theta)$ and $\zeta_{_{\mbox{\footnotesize zf}}}(M, K , \Theta)$) are given by
\begin{eqnarray}
\label{eqn_12_mrc}
\frac{R}{\zeta_{_{\mbox{\footnotesize mrc}}}(M, K , \Theta)}
& \hspace{-3mm} = & {\Big (}   \alpha K \gamma_{\mbox{\footnotesize mrc}} + ( K \rho_d + M \rho_r + \rho_s)  {\Big )} \nonumber \\
& \hspace{-3mm}   =  &    M \rho_r + K \rho_d + \rho_s   \nonumber \\
&  &   + \, \frac{  \alpha K {\Big (} 2^{\frac{R}{K}} - 1 {\Big )}}{ (M - 1) \, - \, (K - 1) (2^{\frac{R}{K}} - 1)}
\end{eqnarray}
and
\begin{eqnarray}
\label{eqn_12_zf}
\frac{R}{\zeta_{_{\mbox{\footnotesize zf}}}(M, K , \Theta)}
& \hspace{-3mm} = & {\Big (}   \alpha K \gamma_{\mbox{\footnotesize zf}} + ( K \rho_d + M \rho_r + \rho_s)  {\Big )} \nonumber \\
& \hspace{-8mm}   =  & \hspace{-4mm} \frac{\alpha K (2^{\frac{R}{K}} - 1)}{(M - K)}   + (M \rho_r + K \rho_d + \rho_s).
\end{eqnarray}
For a given $\Theta = (R, \alpha, \rho_r, \rho_d, \rho_s)$ the optimal energy efficiency $\zeta_{_{\mbox{\footnotesize mrc}}}^{\star}(\Theta)$ and $\zeta_{_{\mbox{\footnotesize zf}}}^{\star}(\Theta)$ are therefore given by
\begin{eqnarray}
\label{zeta_opt_zf_mrc}
\frac{R}{\zeta_{_{\mbox{\footnotesize mrc}}}^{\star}(\Theta)} & = & \min_{(M,K) \, \in \, {\mathcal Z}_{_R} }  \, \frac{R}{\zeta_{_{\mbox{\footnotesize mrc}}}(M, K , \Theta)} \nonumber \\
\frac{R}{\zeta_{_{\mbox{\footnotesize zf}}}^{\star}(\Theta)} & = & \min_{ \substack{ (M,K) \, | \, M, K \, \in {\mathbb Z} \\  M > K \geq 1 }}  \, \frac{R}{\zeta_{_{\mbox{\footnotesize zf}}}(M, K , \Theta)}.
\end{eqnarray}
\section{Energy-Spectral efficiency trade-off of the MRC receiver for large $R$}
\label{largeR_sec}
In \cite{prev_work}, for the ZF receiver we had shown that with a fixed $(\alpha, \rho_r, \rho_d)$, the optimal number of UTs and BS antennas increases with increasing $R$.\footnote{\footnotesize{It is clear from (\ref{zeta_opt_zf_mrc}) that since $\rho_s$ does not depend on $(M,K)$, the optimal $(M,K)$ is independent of $\rho_s$.}} In the regime where $R$ is sufficiently large (i.e., when the optimal number of UTs and BS antennas is much larger than one and two respectively), a tight approximation to the optimal energy efficiency of the ZF receiver was then derived analytically by relaxing the optimization variables $(M,K)$ in (\ref{zeta_opt_zf_mrc}) to be real valued. This approximation denoted by $\zeta_{_{\mbox{\footnotesize zf}}}^{\prime\prime}(\Theta)$ is given by
\begin{eqnarray}
\label{zeta_zf_svar}
\frac{R}{\zeta_{_{\mbox{\footnotesize zf}}}^{\prime\prime}(\Theta)} & = & \hspace{-2mm} \min_{\substack{M, K \, \in {\mathbb R} \\  M > K \,,\, K \geq 1 }}  \, \frac{R}{\zeta_{_{\mbox{\footnotesize zf}}}(M, K , \Theta)} \nonumber \\
& \hspace{-8mm} = &  \min_{\substack{ K \, \in {\mathbb R} \\  K \geq 1 }}  {\Bigg [} \, {\Bigg \{} \min_{\substack{ M \, \in {\mathbb R} \\  M > K }}  \frac{\alpha K (2^{\frac{R}{K}} - 1)}{(M - K)}   +  (M - K) \rho_r  {\Bigg \}} \nonumber \\
& &  \hspace{13mm} \rho_s + K (\rho_r + \rho_d) {\Bigg ]}.\nonumber \\
& \hspace{-8mm} = & \min_{\substack{ K \, \in {\mathbb R} \\  K \geq 1 }}  {\Bigg [} \, {\Bigg \{} \min_{\substack{ t \, \in {\mathbb R} \\  t > 0}}  \frac{\alpha K (2^{\frac{R}{K}} - 1)}{t}   +  t \rho_r  {\Bigg \}} \nonumber \\
& &  \hspace{13mm} \rho_s + K (\rho_r + \rho_d) {\Bigg ]}.\nonumber \\
& \hspace{-19mm} = & \hspace{-12mm} \rho_s + {\Big (} \min_{\substack{ K \, \in {\mathbb R} \\  K \geq 1 }}  \,2 \sqrt{\alpha \rho_r K (2^{R/K} - 1)} + K (\rho_r + \rho_d) {\Big )}.
\end{eqnarray}
In \cite{prev_work} we had considered the special case where $\rho_s = 0$ to understand only the impact of $(\rho_r, \rho_d)$ alone on the optimal energy-spectral efficiency trade-off curve. However, results in \cite{prev_work} can be very simply generalized to the $\rho_s > 0$ case. This is because, the problem of jointly optimizing the energy efficiency w.r.t. $(M,K)$ is independent of $\rho_s$ since $\rho_s$ is a constant which does not depend on $(M,K)$.
The following result from \cite{prev_work} is useful later in this paper to show that the ZF receiver has a strictly better optimal energy efficiency than the MRC receiver in the large $R$ regime. 
\begin{myresult}
\label{result6_gcom14}
[Theorem $7$ in \cite{prev_work}]

For\footnote{\footnotesize{The deterministic function $f(\cdot)$ and the positive scalar $R_c$ have been defined in \cite{prev_work}. $R_c$ depends only on $(\alpha, \rho_r, \rho_d)$ and is therefore fixed.}} $R > f(R_c)$
\begin{eqnarray}
\label{zeta_large_R_eqn_f2}
 ( \rho_r + \rho_d )   < & \frac{\log_2{\Big (} 1 + \frac{R}{R_c} (2^{R_c} - 1) {\Big )}}{  \zeta_{zf}^{\prime}(\Theta)} \,  & <  4 ( \rho_r + \rho_d)
\end{eqnarray}
where $\zeta_{zf}^{\prime}(\Theta)$ is the optimal energy efficiency of the ZF receiver when $\rho_s = 0$.
\end{myresult}

Since $\rho_s$ does not depend on $(M,K)$ we have
\begin{eqnarray}
\label{rel_dprime}
\frac{R}{\zeta_{zf}^{\prime\prime}(\Theta)} & = &  \frac{R}{\zeta_{zf}^{\prime}(\Theta)} + \rho_s.
\end{eqnarray}
Using (\ref{rel_dprime}) in (\ref{zeta_large_R_eqn_f2}) we get for all $R > f(R_c)$
\begin{eqnarray}
\label{result123_1}
\frac{\log_2{\Big (} 1 + \frac{R}{R_c} (2^{R_c} - 1) {\Big )}}{(\rho_r + \rho_d) \, \zeta_{zf}^{\prime\prime}(\Theta)}  & \hspace{-3mm} > & \hspace{-3mm} 1 + \rho_s \frac{\log_2{\Big (} 1 + \frac{R}{R_c} (2^{R_c} - 1) {\Big )}}{R (\rho_r + \rho_d)}  \nonumber \\
\frac{\log_2{\Big (} 1 + \frac{R}{R_c} (2^{R_c} - 1) {\Big )}}{(\rho_r + \rho_d) \, \zeta_{zf}^{\prime\prime}(\Theta)}  & \hspace{-3mm}  < &  \hspace{-3mm}   4 + \rho_s \frac{\log_2{\Big (} 1 + \frac{R}{R_c} (2^{R_c} - 1) {\Big )}}{R (\rho_r + \rho_d)}.  \nonumber \\
\end{eqnarray}
In the large $R$ regime, ${\Big [} \log_2{\Big (} 1 + \frac{R}{R_c} (2^{R_c} - 1) {\Big )} {\Big ]} /R$ decreases with increasing $R$. Let $R_0 > 0$ be defined as the smallest positive number such that for all $R > R_0$
\begin{eqnarray}
\label{result123_2}
\frac{\rho_s}{\rho_r + \rho_d} \, \frac{\log_2{\Big (} 1 + \frac{R}{R_c} (2^{R_c} - 1) {\Big )}}{R} & \hspace{-2mm} < & \hspace{-2mm} 1 \,\,\,,\,\,\, \forall \, R > R_0.
\end{eqnarray}
Combining (\ref{result123_1}) and (\ref{result123_2}) it follows that for all $R > \max(f(R_c), R_0)$
\begin{eqnarray}
\label{zeta_large_R_eqn_f}
 ( \rho_r + \rho_d )   < & \frac{\log_2{\Big (} 1 + \frac{R}{R_c} (2^{R_c} - 1) {\Big )}}{  \zeta_{zf}^{\prime\prime}(\Theta)} \,  & <  5 ( \rho_r + \rho_d).
\end{eqnarray}
In this paper, for the MRC receiver we propose a similar approximation to the optimal energy efficiency (as is done for the ZF receiver in \cite{prev_work}, see (\ref{zeta_zf_svar})) which is given by
\begin{eqnarray}
\label{mrc_zeta_prime_eqn}
\frac{R}{\zeta_{_{\mbox{\footnotesize mrc}}}^{\prime\prime}(\Theta)} & = & \min_{(M,K) \in {\mathcal A}_{_R}}  \, \frac{R}{\zeta_{_{\mbox{\footnotesize mrc}}}(M, K , \Theta)}.
\end{eqnarray}
where
\begin{eqnarray}
\label{arset_def}
{\mathcal A}_{_R} & \Define & {\Big \{ } (M , K)  \,|\, M, K \in {\mathbb R} \,,\, K \geq 1 \,,\, \nonumber \\
& & \hspace{12mm}   (M - 1) > (K - 1) (2^{R/K} - 1)  {\Big \}}.
\end{eqnarray}
Clearly $\zeta_{_{\mbox{\footnotesize mrc}}}^{\prime\prime}(\Theta) > \zeta_{_{\mbox{\footnotesize mrc}}}^{\star}(\Theta)$.
Exhaustive simulations reveal that $\zeta_{_{\mbox{\footnotesize mrc}}}^{\prime\prime}(\Theta) \approx \zeta_{_{\mbox{\footnotesize mrc}}}^{\star}(\Theta)$ for sufficiently large $R$ (see Fig.~\ref{fig_0} in Section \ref{sec_sim}). Let the optimal number of UTs and BS antennas be defined by
\begin{eqnarray}
{\Big (} M_{_{\mbox{\footnotesize mrc}}}^{\star}(\Theta) \,,\, K_{_{\mbox{\footnotesize mrc}}}^{\star}(\Theta) {\Big )} & \Define & \arg \min_{(M,K) \, \in \, {\mathcal Z}_{_R} }  \, \frac{R}{\zeta_{_{\mbox{\footnotesize mrc}}}(M, K , \Theta)} \nonumber \\
\end{eqnarray}
where ${\mathcal Z}_{_R}$ is defined in (\ref{zrset_def}).
Through simulations we have observed that both $M_{_{\mbox{\footnotesize mrc}}}^{\star}(\Theta)$ and $K_{_{\mbox{\footnotesize mrc}}}^{\star}(\Theta)$ increase with increasing $R$. It is also observed that $\zeta_{_{\mbox{\footnotesize mrc}}}^{\prime\prime}(\Theta) \approx \zeta_{_{\mbox{\footnotesize mrc}}}^{\star}(\Theta)$ when $M_{_{\mbox{\footnotesize mrc}}}^{\star}(\Theta) \gg 1$ and $K_{_{\mbox{\footnotesize mrc}}}^{\star}(\Theta) \gg 1$, which happens when $R$ is large (see Fig.~\ref{fig_19} in Section \ref{sec_sim}). For this reason, subsequently in this paper we refer to both $\zeta_{_{\mbox{\footnotesize mrc}}}^{\star}(\Theta)$ and $\zeta_{_{\mbox{\footnotesize mrc}}}^{\prime\prime}(\Theta)$ as the optimal energy efficiency when $R$ is large.
The following theorem reduces (\ref{mrc_zeta_prime_eqn}) to a single variable optimization problem.
 
\begin{mytheorem}
\label{thmthm_gcom}
For any given $\Theta$
\begin{eqnarray}
\label{mrc_prime_single_var}
\frac{R}{\zeta_{_{\mbox{\footnotesize mrc}}}^{\prime\prime}(\Theta) } & =  &  \min_{\substack{K \in {\mathbb R}  \\  K \geq 1 }}  \, g_{_R}(K)  \,\,\,\,,\,\,\,\, \mbox{where} \nonumber \\
g_{_R}(K) & \Define &{\Bigg (}  2 \sqrt{\alpha \rho_r K (2^{R/K} - 1)} + \rho_r  + \rho_s \nonumber \\
& & \hspace{2mm}  + K \rho_d + (K - 1) \rho_r (2^{R/K} - 1) {\Bigg )}.
\end{eqnarray}
\end{mytheorem}

{\em Proof:}
Using (\ref{eqn_12_mrc}) for the expression of $R/\zeta_{_{\mbox{\footnotesize mrc}}}(M, K , \Theta)$ in (\ref{mrc_zeta_prime_eqn}) we get
\begin{eqnarray}
\label{step_a_b_1}
 \min_{(M,K) \in  {\mathcal A}_{_R}}  \frac{R}{\zeta_{_{\mbox{\footnotesize mrc}}}(M, K, \Theta)} &  &   \nonumber \\ 
&  \hspace{-55mm} = &  \hspace{-31mm} \min_{(M,K) \in  {\mathcal A}_{_R} }  \,  {\Bigg [}  {\Big (}  (M - 1) \, - \, (K - 1) (2^{\frac{R}{K}} - 1) {\Big )}\rho_r  \nonumber \\
& &  \hspace{-20mm}   \,\, + \, \frac{  \alpha K {\Big (} 2^{\frac{R}{K}} - 1 {\Big )}}{ (M - 1) \, - \, (K - 1) (2^{\frac{R}{K}} - 1)} \nonumber \\
& &   \hspace{-20mm} \,\, +   \rho_s + \rho_r + K \rho_d + (K - 1) \rho_r (2^{\frac{R}{K}} - 1)  {\Bigg ] }\nonumber \\
& \hspace{-55mm} { (a) \atop = }  & \hspace{-30mm}   \min_{\{ K \in {\mathbb R} \, | \, K \geq 1 \}} \, {\Bigg [} \,\, h(K)  + \rho_r + K \rho_d  + \rho_s \nonumber \\
&  &  \hspace{-7mm} + (K - 1) \rho_r (2^{\frac{R}{K}} - 1) {\Bigg ]} 
\end{eqnarray}
where $h(K)$ is given by
\begin{eqnarray}
\label{step_a_b_hk}
h(K) & \hspace{-5mm} \Define & \hspace{-16mm} \min_{\substack{ M \in {\mathbb R}  \\  (M - 1)  > (K - 1) (2^{R/K} - 1) } }  \hspace{-3mm} {\Bigg (}  {\Big (} (M - 1) \, - \, (K - 1) (2^{\frac{R}{K}} - 1) {\Big )}\rho_r  \nonumber \\
& &  \hspace{15mm} \, + \, \frac{  \alpha K {\Big (} 2^{\frac{R}{K}} - 1 {\Big )}}{ (M - 1) \, - \, (K - 1) (2^{\frac{R}{K}} - 1)} \, {\Bigg )}\nonumber \\
& = & \min_{\substack{ t \in {\mathbb R} \,,\, t  > 0 }} \, {\Bigg [}  t \rho_r  \, + \, \frac{\alpha K {\Big (} 2^{\frac{R}{K}} - 1 {\Big )}}{t}  \, {\Bigg ]} \nonumber \\
& { (a) \atop  = }  & 2 \sqrt{\alpha \rho_r K {\Big (} 2^{\frac{R}{K}} - 1 {\Big )}}.
\end{eqnarray}
Using the expression of $h(K)$ in step (a) of (\ref{step_a_b_hk}) into step (a) of (\ref{step_a_b_1}) we get
(\ref{mrc_prime_single_var}).
$\hfill\blacksquare$

\begin{myremark}
\label{finiteMK_remark}
In (\ref{mrc_prime_single_var}) $g_{_R}(K)$ corresponds to the total system power consumed
for a given $R$ and $K$ ($M$ being chosen optimally for the given $(R,K)$, see (\ref{step_a_b_hk})).
In the R.H.S. of the expression for $g_{_R}(K)$ in (\ref{mrc_prime_single_var}), the term $K \rho_d$ corresponds to the power consumed by the multiuser signal processing at the BS and by the transmitter circuitry (except PA) in the UTs. With fixed $\Theta$ and increasing $K$, it is clear that this term increases. At the same time, with increasing $K$ the per-user information rate $R/K$ decreases (since $\Theta$ and therefore $R$ is fixed) which results in a decrease in the required power to be radiated by each UT. Therefore, with increasing $K$ the power consumed by the PAs in all the UTs decreases (corresponds to the term $2 \sqrt{\alpha \rho_r K (2^{R/K} - 1)}$ in (\ref{mrc_prime_single_var})).
Hence, there is a trade-off involved in choosing the optimal $K$ which minimizes the total system power consumed. Therefore for a fixed $\Theta$, the optimal $K$ which maximizes the energy efficiency is finite.

Further from the minimization in (\ref{step_a_b_hk}), it is clear that for a given $(\Theta,K)$, the optimal number of BS antennas is given by
\begin{eqnarray}
\label{mtheta_def}
M_{_{\mbox{\footnotesize mrc}}}^{\prime\prime}(\Theta, K) & \Define & \arg  \hspace{-3mm} \min_{ \{ M \, | \, (M,K) \in {\mathcal A}_{_R} \}} \, \frac{R}{\zeta_{_{\mbox{\footnotesize mrc}}}(M,K,\Theta)} \nonumber \\
& \hspace{-29mm}  = & \hspace{-16mm} 1 + (K - 1) {\Big (} 2^{\frac{R}{K}} - 1 {\Big )} \, + \, \sqrt{\frac{\alpha  K {\Big (} 2^{\frac{R}{K}} - 1 {\Big )}}{\rho_r}}
\end{eqnarray}
which implies that the optimal number of BS antennas is also finite (since the optimal $K$ is finite).
This result on the finiteness of the optimal $(M,K)$ is in contrast to the scenario where transceiver power consumption is not taken into consideration (i.e., $\rho_r = \rho_d = \rho_s =  0$), due to which the energy efficiency grows unbounded with increasing $(M,K)$, and hence the optimal $(M,K)$ is not finite \cite{HNQ}.
\end{myremark}

\begin{myremark}
\label{rhor_remark}
From the expression of $g_{_R}(K)$ in (\ref{mrc_prime_single_var}) it is clear that for a fixed $(R,K)$, $g_{_R}(K)$ increases with increasing $\rho_r$, $\rho_d$ and $\rho_s$. This therefore implies that for a fixed $(\alpha,R)$, $\zeta_{_{\mbox{\footnotesize mrc}}}^{\prime\prime}(\Theta)$ decreases with increasing $\rho_r$, $\rho_d$ and $\rho_s$. This conclusion is also intuitive since any increase in $\rho_r$, $\rho_d$ and $\rho_s$ 
will increase the total system power consumed, thereby decreasing the energy efficiency (since $R$ is fixed). 
\end{myremark}

\begin{myremark}
In \cite{HNQ}, for the scenario $\rho_r = \rho_d = \rho_s = 0$ it has been shown that for a fixed $(M,K)$ and sufficiently large $R$, the energy efficiency of both the ZF and the MRC receiver decreases with increasing $R$ and are asymptotically zero as $R \rightarrow \infty$.
For the scenario ($\rho_r \ne 0 \,,\, \rho_d \ne 0  \,,\, \rho_s \ne 0$) considered in this paper, the following corollary to Theorem \ref{thmthm_gcom} shows that for the MRC receiver, by increasing $(M,K)$ appropriately with increasing $R$, the energy efficiency will asymptotically converge to a {\em positive constant} as $R \rightarrow \infty$.
\end{myremark}

\begin{corollary}
\label{cor_largeRR}
For a fixed $(\alpha, \rho_r, \rho_d, \rho_s)$ and increasing $R$, let $({\Tilde M(R)}, {\Tilde K(R) })$ be given by
\begin{eqnarray}
\label{tildeMK_def}
{\Tilde K(R)} & \Define & \frac{R}{c} \,\,,\,\,
{\Tilde M(R)} \, \Define \, M_{_{\mbox{\footnotesize mrc}}}^{\prime\prime}(\Theta, {\Tilde K(R)}) 
\end{eqnarray}
where the function $M_{_{\mbox{\footnotesize mrc}}}^{\prime\prime}(\cdot, \cdot)$ is defined in (\ref{mtheta_def}), and $c > 0$ is a constant.
Then for $R > c$
\begin{eqnarray}
\label{prf_trv}
\zeta_{_{\mbox{\footnotesize mrc}}}({\Tilde M(R)}, {\Tilde K(R)}, \Theta)  & \leq &  \zeta_{_{\mbox{\footnotesize mrc}}}^{\prime\prime}(\Theta)
\end{eqnarray}
where $\zeta_{_{\mbox{\footnotesize mrc}}}(\cdot, \cdot, \cdot)$ is defined in (\ref{eqn_12_mrc}).
Also for any $c > 0$
\begin{eqnarray}
\label{prf_diffc}
\lim_{R \rightarrow \infty} \, \zeta_{_{\mbox{\footnotesize mrc}}}^{\prime\prime}(\Theta)  & \geq &  \frac{c}{\rho_d + \rho_r (2^c - 1) } \, > \, 0.
\end{eqnarray}
\end{corollary}

{\em Proof}:
For $R > c$, ${\Tilde K(R)} = (R/c)  > 1$ and hence $({\Tilde M(R)}, {\Tilde K(R)}) \in {\mathcal A}_{_R}$.\footnote{\footnotesize{From (\ref{mtheta_def}) it follows that $({\Tilde M(R)}  - 1) - ({\Tilde K(R)} - 1) (2^{R/{\Tilde K(R)}} - 1)  =  \sqrt{\alpha  {\Tilde K(R)} {\Big (} 2^{R/{\Tilde K(R)}} - 1 {\Big )} / \rho_r } \, > \, 0$.}} Therefore the inequality in (\ref{prf_trv}) now follows clearly from the definition of $\zeta_{_{\mbox{\footnotesize mrc}}}^{\prime\prime}(\Theta)$ in (\ref{mrc_zeta_prime_eqn}). Using (\ref{eqn_12_mrc}) with $(M,K) = ({\Tilde M(R)}, {\Tilde K(R)})$ we get
\begin{eqnarray}
\label{RR_eqn}
\zeta_{_{\mbox{\footnotesize mrc}}}({\Tilde M(R)},  {\Tilde K(R)}, \Theta)   & = &  \nonumber \\
& \hspace{-68mm} &  \hspace{-38mm } \frac{1}{ {\Bigg (}  2 \sqrt{\frac{\alpha \rho_r}{R} \frac{(2^c - 1)}{c} }  + \frac{\rho_r}{R}  +  \frac{\rho_s}{R} + \frac{\rho_d}{c} + \rho_r {\Big (} \frac{1}{c} - \frac{1}{R}  {\Big )} (2^c - 1) {\Bigg )} }. \nonumber \\
\end{eqnarray}
Taking the limit $R \rightarrow \infty$ on both sides of (\ref{RR_eqn}) we get
\begin{eqnarray}
\label{Rvar_eff}
\lim_{R \rightarrow \infty} \, \zeta_{_{\mbox{\footnotesize mrc}}}({\Tilde M(R)},  {\Tilde K(R)}, \Theta)  & = &   \frac{c}{\rho_d + \rho_r (2^c - 1) }.
\end{eqnarray}
Using (\ref{Rvar_eff}) and (\ref{prf_trv}) we finally get (\ref{prf_diffc}).
$\hfill\blacksquare$

\begin{myremark}
\label{remark_77}
From (\ref{tildeMK_def}) and (\ref{mtheta_def}) it can be seen that both $ {\Tilde M(R)}$ and ${\Tilde K(R)}$ increase with increasing $R$. Since ${\Tilde K(R)}$ increases linearly with $R$, it follows that with increasing $R$ the per-user spectral efficiency is constant (i.e., $R/{\Tilde K(R)} = c$).
Therefore, (\ref{Rvar_eff}) in Corollary \ref{cor_largeRR} shows that with increasing $R$, it is possible to increase $(M,K) = ({\Tilde M(R)}, {\Tilde K(R)})$ in such a way that the energy efficiency converges to a positive constant as $R \rightarrow \infty$ (while maintaining a constant per-user spectral efficiency).
From the denominator of the R.H.S. of (\ref{RR_eqn}) it is clear that by choosing $(M,K) = ({\Tilde M(R)}, {\Tilde K(R)})$, as $R \rightarrow \infty$ the total system power consumption is dominated by the power consumed by the transmitter circuitry at the UTs/multiuser processing at the BS (corresponding to the term $\rho_d/c$) and the power consumed by the RF circuits/baseband hardware in the BS (corresponding to the term $\rho_r (2^c - 1)/c$).\footnote{\footnotesize{In the denominator of the R.H.S. of (\ref{RR_eqn}) the term $\rho_s/R$ corresponds to the power consumption at the BS due to operations which are independent of the number of BS antennas and the number of users. It is clear that $\rho_s/R$ has little impact on $\zeta_{_{\mbox{\footnotesize mrc}}}({\Tilde M(R)},  {\Tilde K(R)}, \Theta)$ when $R$ is large.}}

Corollary \ref{cor_largeRR} however does not tell us about the behaviour of the optimal energy efficiency $\zeta_{_{\mbox{\footnotesize mrc}}}^{\star}(\Theta)$ with increasing $R$. It does not tell us whether the optimal energy efficiency also converges to a constant just like $\zeta_{_{\mbox{\footnotesize mrc}}}({\Tilde M(R)},  {\Tilde K(R)}, \Theta) $ or does it increase unbounded with increasing $R$ (as is the case in ZF receivers). Nevertheless, through exhaustive simulations we have seen that the optimal energy efficiency of the MRC receiver also {\em converges to a constant} with increasing $R$ (see Fig.~\ref{fig_0} in Section \ref{sec_sim}). As conjectured in Remark \ref{remark_77}, exhaustive simulations have also revealed that with the optimal choice of UTs and BS antennas, i.e., $(M,K) = (M_{_{\mbox{\footnotesize mrc}}}^{\star}(\Theta), K_{_{\mbox{\footnotesize mrc}}}^{\star}(\Theta))$
most of the total system power consumption is attributed to the power consumed by the transmitter circuitry in the UTs (except the PA) and by the BS (see Fig.~\ref{fig_13} in Section \ref{sec_sim}).
\end{myremark}

The following theorem will be used later to show that the ZF receiver is more energy efficient than the MRC receiver when $R$ is sufficiently large.
\begin{mytheorem}
\label{theorem3_gcom14}
For any given $\Theta$ satisfying
\begin{eqnarray}
\label{theorem_conditions}
R  & >  &  \max(R_1 \,,\, R_2)    \,\,\,\,,\,\,\, \mbox{where} \nonumber \\
R_1 & \Define & \max{\Bigg (}4 \,,\, 4 \log_2{\Big (} 1 +  \frac{\alpha}{\rho_r}{\Big )} {\Bigg )}  \,\,\,\mbox{and} \nonumber \\
R_2 & \Define &  \max{\Bigg (}  \log_2 {\Big (} 1 + \frac{9 \rho_d^2}{\alpha \rho_r} {\Big )}  \,,\, 2 \log_2{\Big (}   \frac{49 \rho_r}{\alpha} {\Big )}  {\Bigg )}
\end{eqnarray}
i.e., sufficiently large $R$, it follows that
\begin{eqnarray}
\label{ref_thm222}
\zeta_{_{\mbox{\footnotesize mrc}}}^{\prime\prime}(\Theta) & < & \frac{1}{\min{\Bigg (} \frac{1}{\zeta_{_{\mbox{\footnotesize zf}}}^{\prime\prime}(\Theta) }   \,,\, \rho_d + \frac{\rho_r}{R} +  \frac{\rho_s}{R}  {\Bigg )}}.
\end{eqnarray}
\end{mytheorem}

{\em Proof:}
In (\ref{mrc_prime_single_var}) separating the range of the optimization variable $K$ into
intervals\footnote{\footnotesize{We make use of the fact that $R > 1$, as stated in (\ref{theorem_conditions}).}} $[ 1  \,,\, R)$ and $[R \,,\, \infty)$ we get
\begin{eqnarray}
\label{prf_2_eqn_1}
\frac{R}{\zeta_{_{\mbox{\footnotesize mrc}}}^{\prime\prime}(\Theta) } 
& = & \min{\Bigg (}  \, {\Big (} \, \min_{\substack{K \in {\mathbb R} \\ R > K \geq 1 }}  \, g_{_R}(K)  {\Big )}  \,,\,  {\Big (} \, \min_{\substack{K \in {\mathbb R} \\ K \geq R }}  \, g_{_R}(K)  {\Big )} \,  {\Bigg )}. \nonumber \\
\end{eqnarray}
From (\ref{mrc_prime_single_var}) we note that $g_{_R}(K) > ( \rho_r + \rho_s +  K \rho_d)$ for any $K > 1$. If $K \geq R$, it follows that
$K > 1$ also, since $R > 1$ (see (\ref{theorem_conditions})). Using these facts we get
\begin{eqnarray}
\label{ineq_1_prf}
\min_{\substack{K \in {\mathbb R} \\ K \geq R }}  g_{_R}(K) & > & \min_{\substack{K \in {\mathbb R} \\ K \geq R }} (\rho_r + \rho_s + K \rho_d ) \nonumber \\
& = & \rho_r +  \rho_s + R \rho_d.
\end{eqnarray}
From the expression of $g_{_R}(K)$ in (\ref{mrc_prime_single_var}) we also have
\begin{eqnarray}
\label{grk_new}
g_{_R}(K) & = & {\Bigg (}  2 \sqrt{\alpha \rho_r K (2^{R/K} - 1)} + K (\rho_r + \rho_d)  + \rho_s \nonumber \\
& & \hspace{6mm} + (K - 1) \rho_r (2^{R/K} - 2) {\Bigg )}.
\end{eqnarray}
For any $R$ satisfying the conditions in (\ref{theorem_conditions}), simple algebraic manipulations
show that $g_{_R}(1)   >   g_{_R}(4)$.\footnote{\footnotesize{From the conditions in (\ref{theorem_conditions}) it follows that $\sqrt{\alpha \rho_r (2^R - 1)} > 3 \rho_d$, $\rho_r (2^{R/4} - 1)  > \sqrt{\alpha \rho_r (2^{R/4} - 1)}$ and $\sqrt{\alpha \rho_r (2^R - 1)} > 7 \rho_r (2^{R/4} - 1)$. Using these three inequalities it can be shown that $g_{_R}(1)   >  g_{_R}(4)$.}}
From (\ref{theorem_conditions}) we have $R > 4$ and therefore $4 \in [1 \,,\, R)$. Since $g_{_R}(1)   >   g_{_R}(4)$, $K=1$ is not a minimum of $g_{_R}(K)$ when $K \in [1 \,,\, R)$, i.e.
\begin{eqnarray}
\label{step_1_2_pr}
\min_{\substack{K \in {\mathbb R} \\ R > K \geq 1 }}  \, g_{_R}(K)  & = &   \min_{\substack{K \in {\mathbb R} \\ R > K > 1 }}  \, g_{_R}(K).
\end{eqnarray}
From (\ref{step_1_2_pr}) we have
\begin{eqnarray}
\label{ineq_2_prf}
\min_{\substack{K \in {\mathbb R} \\ R > K \geq 1 }}  \, g_{_R}(K)  & { = } &  \min_{\substack{K \in {\mathbb R} \\ R > K >  1 }}  \, g_{_R}(K)  \nonumber \\
&  \hspace{-26mm } {(a) \atop  = } &  \hspace{-14mm} \min_{\substack{K \in {\mathbb R} \\ R > K > 1 }}  {\Bigg (} 2 \sqrt{\alpha \rho_r K (2^{R/K} - 1)} + K (\rho_r + \rho_d) \nonumber \\
& & + \, \rho_s + (K - 1) \rho_r (2^{R/K} - 2) {\Bigg )} \nonumber \\
& \hspace{-26mm }   {(b) \atop > }  &  \hspace{-14mm} \min_{\substack{K \in {\mathbb R} \\  R >  K > 1 }}  \,\, 2 \sqrt{\alpha \rho_r K (2^{R/K} - 1)} + K (\rho_r + \rho_d)  + \rho_s \nonumber \\
& \hspace{-26mm } \geq   & \hspace{-14mm}  \min_{\substack{K \in {\mathbb R} \\  K \geq 1 }}  \,\, 2 \sqrt{\alpha \rho_r K (2^{R/K} - 1)} + K (\rho_r + \rho_d)  + \rho_s \nonumber \\
& \hspace{-26mm }  {(c) \atop = } & \hspace{-14mm}   \frac{R}{\zeta_{_{\mbox{\footnotesize zf}}}^{\prime\prime}(\Theta) }
\end{eqnarray}
where step (a) follows from (\ref{grk_new}) and step (b) follows from the fact
that for $1 < K < R$, $(K - 1) (2^{R/K} - 2)  > 0$.
Step (c) follows from (\ref{zeta_zf_svar}).
Using (\ref{ineq_1_prf}) and (\ref{ineq_2_prf}) in (\ref{prf_2_eqn_1}) we get (\ref{ref_thm222}).
$\hfill\blacksquare$
 
From (\ref{zeta_large_R_eqn_f}) we know that $\zeta_{_{\mbox{\footnotesize zf}}}^{\prime\prime}(\Theta) \propto O(\log(R))$ for sufficiently large $R$. Hence $(1/\zeta_{_{\mbox{\footnotesize zf}}}^{\prime\prime}(\Theta) ) < \rho_d + (\rho_r/R) + (\rho_s/R)$ for sufficiently large $R$. Using this fact in (\ref{ref_thm222}) leads us to the following Corollary.
\begin{corollary}
\label{cor_22}
For any given $\Theta$ with $R$ sufficiently large
\begin{eqnarray}
\label{cor_conds}
R &  \hspace{-3mm } > &  \hspace{-3mm} \max {\Bigg (}  R_0, R_1, R_2, f(R_c) \,,\, \frac{R_c \, {\Big (}  32^{\frac{(\rho_r + \rho_d)}{\rho_d}}  - 1 {\Big )}}{(2^{R_c} - 1)} {\Bigg )}
\end{eqnarray}
it holds that
\begin{eqnarray}
\label{cor_ineq}
\zeta_{_{\mbox{\footnotesize mrc}}}^{\prime\prime}(\Theta) & < & \zeta_{_{\mbox{\footnotesize zf}}}^{\prime\prime}(\Theta)
\end{eqnarray}
where $R_1,R_2$ are defined in (\ref{theorem_conditions}), $R_0$ is defined in (\ref{result123_2}) and $f(\cdot), R_c$ are defined in Result \ref{result6_gcom14}.
\end{corollary}

{\em Proof}:
Since $R > \max(R_0 ,f(R_c))$, from (\ref{zeta_large_R_eqn_f}) we have
\begin{eqnarray}
\label{cor4_1}
\frac{1}{\zeta_{_{\mbox{\footnotesize zf}}}^{\prime\prime}(\Theta)} & < &  \frac{5 (\rho_r + \rho_d) }{\log_2{\Big (} 1 + \frac{R}{R_c} (2^{R_c} - 1) {\Big )}}.
\end{eqnarray}
Further since $R > \frac{R_c \, {\Big (}  32^{\frac{(\rho_r + \rho_d)}{\rho_d}}  - 1 {\Big )}}{(2^{R_c} - 1)}$  (see (\ref{cor_conds})), using simple algebraic manipulations we get
\begin{eqnarray}
\label{cor4_2}
 \frac{5 (\rho_r + \rho_d) }{\log_2{\Big (} 1 + \frac{R}{R_c} (2^{R_c} - 1) {\Big )}} & < & \rho_d 
\ <  \rho_d + \frac{\rho_r}{R} + \frac{\rho_s}{R}.
\end{eqnarray}
Combining (\ref{cor4_1}) and (\ref{cor4_2}) we get
$(1/\zeta_{_{\mbox{\footnotesize zf}}}^{\prime\prime}(\Theta)) <    \rho_d + \frac{\rho_r}{R} + \frac{\rho_s}{R}$
for all $R$ satisfying (\ref{cor_conds}).
Using this fact along with Theorem \ref{theorem3_gcom14} completes the proof.
$\hfill\blacksquare$

\begin{myremark}
\label{remark_23}
In the following we explain the observation made in Corollary \ref{cor_22}.
Comparing (\ref{eqn_12_mrc}) and (\ref{eqn_12_zf}) for the same $(M,K,\Theta)$ we see that the only difference in the energy efficiency of the ZF and the MRC receivers is due to the difference in the power consumed by the PAs in the UTs. In the large $R$ regime, for a fixed $(R,M,K)$ the power required to be radiated by each UT is higher in the case of the MRC receiver, since unlike the ZF receiver, the MRC receiver does not cancel multiuser interference (which reduces its post-combining signal-to-noise-and-interference ratio). Comparing (\ref{eqn_12_mrc}) and (\ref{eqn_12_zf}) it is clear that for the same $(R,M,K)$  the ZF receiver has an array gain higher than that of the MRC receiver by $(K - 1)(2^{R/K} - 2)$. For the MRC receiver to have the same array gain as that of the ZF receiver, one possibility is to increase $M$ by $(K - 1)(2^{R/K} - 2)$ in the case of the MRC receiver. However this will lead to an increase in the power consumed by the BS hardware (i.e., see the $M \rho_r$ term in (\ref{eqn_12_mrc})).

With sufficiently large $R$, this extra increase of $(K - 1)(2^{R/K} - 2) \rho_r$ in the power consumed by the BS hardware can be seen as the last term in the expression for the total system power consumed in the R.H.S. of (\ref{grk_new}). 
From (\ref{zeta_zf_svar}) it is clear that the sum of the first three terms in the R.H.S. of (\ref{grk_new}) corresponds
to the total system power consumed when a ZF receiver is used at the BS.
\end{myremark}
\section{Numerical results}
\label{sec_sim}
\begin{figure}[t]
\begin{center}
\epsfig{file=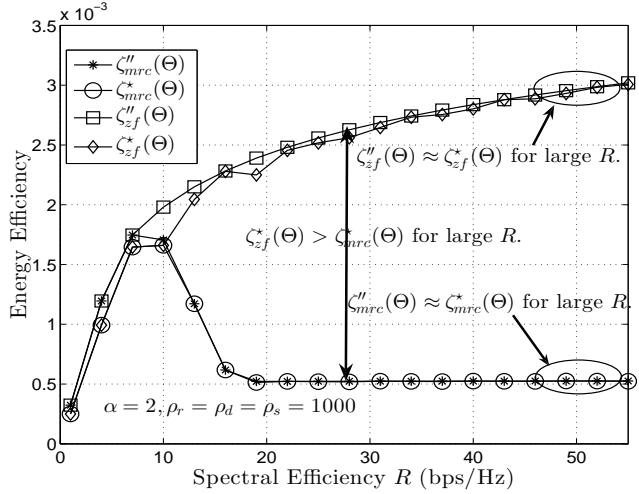, width=90mm,height=68mm}
\end{center}
\vspace{-3mm}
\caption{Optimal energy efficiency versus spectral energy efficiency $R$, for a fixed $\alpha = 2$ and $\rho_r = \rho_d = \rho_s = 10^3$. Note that the ZF receiver has a much better energy efficiency than the MRC receiver when $R$ is large. Further the approximation $\zeta_{_{\mbox{\footnotesize mrc}}}^{\prime\prime}(\Theta) \approx \zeta_{_{\mbox{\footnotesize mrc}}}^{\star}(\Theta)$ is tight for large $R$. Also, the optimal energy efficiency of the MRC receiver converges to a constant as $R \rightarrow \infty$.}
\label{fig_0}
\vspace{-2mm}
\end{figure}
In Fig.~\ref{fig_0} we numerically compute and plot the energy efficiency as a function of increasing spectral efficiency for fixed $\alpha=2$ and $\rho_r = \rho_d = \rho_s = 10^3$. The energy efficiency of both the MRC and the ZF receivers is plotted. We plot both the optimal energy efficiency ($\zeta_{_{\mbox{\footnotesize mrc}}}^{\star}(\Theta) , \zeta_{_{\mbox{\footnotesize zf}}}^{\star}(\Theta)$) and its approximation ($\zeta_{_{\mbox{\footnotesize mrc}}}^{\prime\prime}(\Theta) , \zeta_{_{\mbox{\footnotesize zf}}}^{\prime\prime}(\Theta)$). As stated earlier, it is observed that the approximation to the optimal energy efficiency is tight at sufficiently large $R$ for both the MRC and the ZF receivers. It is also observed that the optimal energy efficiency
of the MRC receiver is strictly less than that of the ZF receiver for large $R$. This confirms the conclusion made in Corollary \ref{cor_22} (see (\ref{cor_ineq})) and Remark \ref{remark_23}.
From Fig.~\ref{fig_0} we also observe that for a fixed $(\alpha, \rho_r, \rho_d, \rho_s)$ the optimal energy efficiency of the MRC receiver converges to a constant value as $R \rightarrow \infty$ (see the paragraph after Remark \ref{remark_77} in Section \ref{largeR_sec}).

\begin{figure}[t]
\begin{center}
\epsfig{file=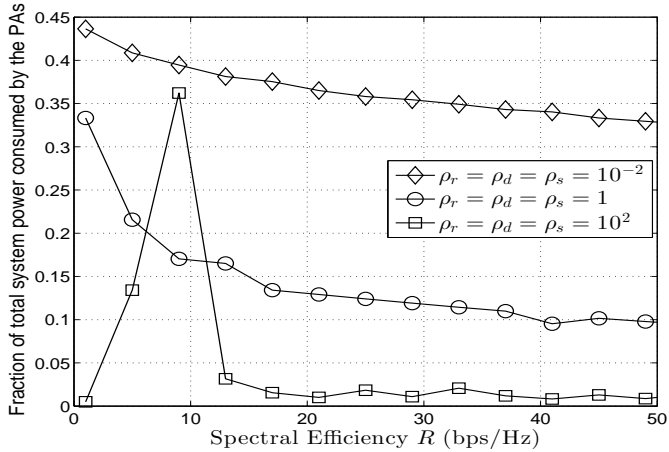, width=90mm,height=60mm}
\end{center}
\vspace{-3mm}
\caption{Fraction of the total system power consumed by the PAs in the UTs versus spectral efficiency $R$ for a fixed $(\alpha,\rho_r, \rho_d, \rho_s)$. $\alpha=2$ and $\rho_r = \rho_d = \rho_s = 10^{-2}, 1, 10^{2}$. In the large $R$ regime, the total system power consumption is dominated by the power consumed in the BS and the transmitter circuitry in the UTs (except the PA).}
\label{fig_13}
\vspace{-3mm}
\end{figure}

In Fig.~\ref{fig_13} we plot the fraction of the total system power consumed by the PAs in the UTs as a function of increasing $R$ for a fixed $\rho_r = \rho_d = \rho_s$ and $\alpha = 2$. For each $R$, we numerically compute the optimal energy efficiency $\zeta_{_{\mbox{\footnotesize mrc}}}^{\star}(\Theta)$ and the corresponding optimal $(M,K)$. With this optimal $(M,K) = (M_{_{\mbox{\footnotesize mrc}}}^{\star}(\Theta) \,,\, K_{_{\mbox{\footnotesize mrc}}}^{\star}(\Theta))$, the fraction of the total system power consumed by the PAs in the UTs is given by $\frac {K  \alpha \gamma_{mrc}}{ K  \alpha \gamma_{mrc} + M \rho_r + K \rho_d  + \rho_s}$ (follows from (\ref{eqn_12_mrc})).
In Fig.~\ref{fig_13} it is seen that with increasing $R$ most of the total system power is consumed by the transmitter circuitry in the UTs and the BS. This shows that in the large $R$ regime, the optimal energy efficiency is {\em limited} by the power consumed in the BS and the transmitter circuitry in the UTs (except the PAs).
This observation is also conjectured in Remark \ref{remark_77} in Section \ref{largeR_sec}.

\begin{figure}[t]
\begin{center}
\epsfig{file=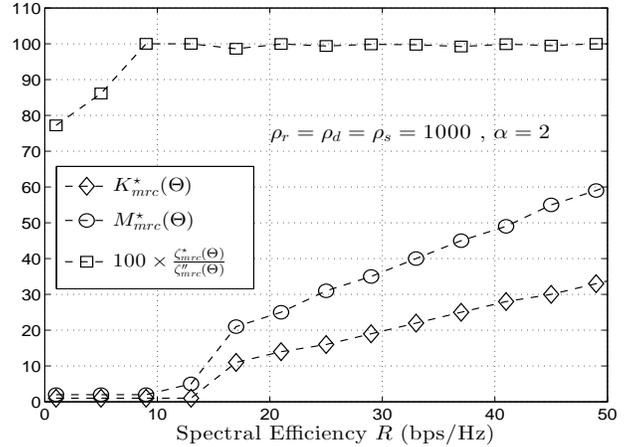, width=90mm,height=64mm}
\end{center}
\vspace{-3mm}
\caption{${\Big (} M_{_{\mbox{\footnotesize mrc}}}^{\star}(\Theta) \,,\, K_{_{\mbox{\footnotesize mrc}}}^{\star}(\Theta) {\Big )}$ versus spectral efficiency $R$ for a fixed $(\alpha, \rho_r,\rho_d, \rho_s) = (2, 10^3, 10^3, 10^3)$. The optimal number of UTs and BS antennas increases with increasing $R$. The approximation $\zeta_{_{\mbox{\footnotesize mrc}}}^{\star}(\Theta) \approx \zeta_{_{\mbox{\footnotesize mrc}}}^{\prime\prime}(\Theta)$ is tight when $M_{_{\mbox{\footnotesize mrc}}}^{\star}(\Theta)  \gg 1$ and $K_{_{\mbox{\footnotesize mrc}}}^{\star}(\Theta) \gg 1$ which happens when $R$ is sufficiently large.}
\label{fig_19}
\vspace{-3mm}
\end{figure}
In Fig.~\ref{fig_19} we plot the optimal number of UTs $K_{_{\mbox{\footnotesize mrc}}}^{\star}(\Theta)$
and the optimal number of BS antennas $M_{_{\mbox{\footnotesize mrc}}}^{\star}(\Theta)$ as a function of increasing $R$ for a fixed $\rho_r = \rho_d = \rho_s = 10^3$ and a fixed $\alpha = 2$. It is observed that the optimal number of UTs and BS antennas increases with $R$. We also plot the ratio between the optimal energy efficiency and its approximation (i.e., $\zeta_{_{\mbox{\footnotesize mrc}}}^{\star}(\Theta)/\zeta_{_{\mbox{\footnotesize mrc}}}^{\prime\prime}(\Theta)$). It is observed that the approximation to the optimal energy efficiency is tight when $K_{_{\mbox{\footnotesize mrc}}}^{\star}(\Theta) \gg 1$ and $M_{_{\mbox{\footnotesize mrc}}}^{\star}(\Theta) \gg 1$.


\end{document}